\documentclass[iop]{emulateapj}
\usepackage{gensymb}

\RequirePackage{lineno}
\PassOptionsToPackage{hyphens}{url}
\usepackage[hidelinks]{hyperref}
\usepackage{amsmath}
\newcommand{\Fermi}{{\it Fermi }}
\newcommand{\fermi}{{\it Fermi}}
\usepackage{nicefrac}
\begin{document}

\title{A Search for Pulsations from Geminga Above 100 GeV with VERITAS}

\author{
E.~Aliu\altaffilmark{1,29},
S.~Archambault\altaffilmark{2},
A.~Archer\altaffilmark{3},
T.~Aune\altaffilmark{4},
A.~Barnacka\altaffilmark{5},
M.~Beilicke\altaffilmark{3},
W.~Benbow\altaffilmark{6},
R.~Bird\altaffilmark{7},
J.~H.~Buckley\altaffilmark{3},
V.~Bugaev\altaffilmark{3},
K.~Byrum\altaffilmark{8},
J.~V~Cardenzana\altaffilmark{9},
M.~Cerruti\altaffilmark{6},
X.~Chen\altaffilmark{10,11},
L.~Ciupik\altaffilmark{12},
M.~P.~Connolly\altaffilmark{13},
W.~Cui\altaffilmark{14},
H.~J.~Dickinson\altaffilmark{9},
J.~Dumm\altaffilmark{15},
J.~D.~Eisch\altaffilmark{9},
M.~Errando\altaffilmark{1},
A.~Falcone\altaffilmark{16},
Q.~Feng\altaffilmark{14},
J.~P.~Finley\altaffilmark{14},
H.~Fleischhack\altaffilmark{11},
P.~Fortin\altaffilmark{6},
L.~Fortson\altaffilmark{15},
A.~Furniss\altaffilmark{17},
G.~H.~Gillanders\altaffilmark{13},
S.~Griffin\altaffilmark{2},
S.~T.~Griffiths\altaffilmark{18},
J.~Grube\altaffilmark{12},
G.~Gyuk\altaffilmark{12},
N.~H{\aa}kansson\altaffilmark{10},
D.~Hanna\altaffilmark{2},
J.~Holder\altaffilmark{19},
T.~B.~Humensky\altaffilmark{20},
C.~A.~Johnson\altaffilmark{17},
P.~Kaaret\altaffilmark{18},
P.~Kar\altaffilmark{21},
M.~Kertzman\altaffilmark{22},
D.~Kieda\altaffilmark{21},
F.~Krennrich\altaffilmark{9},
S.~Kumar\altaffilmark{19},
M.~J.~Lang\altaffilmark{13},
M.~Lyutikov\altaffilmark{14},
A.~S~Madhavan\altaffilmark{9},
G.~Maier\altaffilmark{11},
S.~McArthur\altaffilmark{23},
A.~McCann\altaffilmark{24,a},
K.~Meagher\altaffilmark{25},
J.~Millis\altaffilmark{26},
P.~Moriarty\altaffilmark{13},
R.~Mukherjee\altaffilmark{1},
D.~Nieto\altaffilmark{20},
A.~O'Faol\'{a}in de Bhr\'{o}ithe\altaffilmark{11},
R.~A.~Ong\altaffilmark{4},
A.~N.~Otte\altaffilmark{25},
N.~Park\altaffilmark{23},
M.~Pohl\altaffilmark{10,11},
A.~Popkow\altaffilmark{4},
H.~Prokoph\altaffilmark{11},
E.~Pueschel\altaffilmark{7},
J.~Quinn\altaffilmark{7},
K.~Ragan\altaffilmark{2},
L.~C.~Reyes\altaffilmark{27},
P.~T.~Reynolds\altaffilmark{28},
G.~T.~Richards\altaffilmark{25,b},
E.~Roache\altaffilmark{6},
M.~Santander\altaffilmark{1},
G.~H.~Sembroski\altaffilmark{14},
K.~Shahinyan\altaffilmark{15},
A.~W.~Smith\altaffilmark{21},
D.~Staszak\altaffilmark{2},
I.~Telezhinsky\altaffilmark{10,11},
J.~V.~Tucci\altaffilmark{14},
J.~Tyler\altaffilmark{2},
A.~Varlotta\altaffilmark{14},
S.~Vincent\altaffilmark{11},
S.~P.~Wakely\altaffilmark{23},
A.~Weinstein\altaffilmark{9},
D.~A.~Williams\altaffilmark{17},
A.~Zajczyk\altaffilmark{3},
B.~Zitzer\altaffilmark{8}
}

\altaffiltext{1}{Department of Physics and Astronomy, Barnard College, Columbia University, NY 10027, USA}
\altaffiltext{2}{Physics Department, McGill University, Montreal, QC H3A 2T8, Canada}
\altaffiltext{3}{Department of Physics, Washington University, St. Louis, MO 63130, USA}
\altaffiltext{4}{Department of Physics and Astronomy, University of California, Los Angeles, CA 90095, USA}
\altaffiltext{5}{Harvard-Smithsonian Center for Astrophysics, 60 Garden Street, Cambridge, MA 02138, USA}
\altaffiltext{6}{Fred Lawrence Whipple Observatory, Harvard-Smithsonian Center for Astrophysics, Amado, AZ 85645, USA}
\altaffiltext{7}{School of Physics, University College Dublin, Belfield, Dublin 4, Ireland}
\altaffiltext{8}{Argonne National Laboratory, 9700 S. Cass Avenue, Argonne, IL 60439, USA}
\altaffiltext{9}{Department of Physics and Astronomy, Iowa State University, Ames, IA 50011, USA}
\altaffiltext{10}{Institute of Physics and Astronomy, University of Potsdam, 14476 Potsdam-Golm, Germany}
\altaffiltext{11}{DESY, Platanenallee 6, 15738 Zeuthen, Germany}
\altaffiltext{12}{Astronomy Department, Adler Planetarium and Astronomy Museum, Chicago, IL 60605, USA}
\altaffiltext{13}{School of Physics, National University of Ireland Galway, University Road, Galway, Ireland}
\altaffiltext{14}{Department of Physics and Astronomy, Purdue University, West Lafayette, IN 47907, USA}
\altaffiltext{15}{School of Physics and Astronomy, University of Minnesota, Minneapolis, MN 55455, USA}
\altaffiltext{16}{Department of Astronomy and Astrophysics, 525 Davey Lab, Pennsylvania State University, University Park, PA 16802, USA}
\altaffiltext{17}{Santa Cruz Institute for Particle Physics and Department of Physics, University of California, Santa Cruz, CA 95064, USA}
\altaffiltext{18}{Department of Physics and Astronomy, University of Iowa, Van Allen Hall, Iowa City, IA 52242, USA}
\altaffiltext{19}{Department of Physics and Astronomy and the Bartol Research Institute, University of Delaware, Newark, DE 19716, USA}
\altaffiltext{20}{Physics Department, Columbia University, New York, NY 10027, USA}
\altaffiltext{21}{Department of Physics and Astronomy, University of Utah, Salt Lake City, UT 84112, USA}
\altaffiltext{22}{Department of Physics and Astronomy, DePauw University, Greencastle, IN 46135-0037, USA}
\altaffiltext{23}{Enrico Fermi Institute, University of Chicago, Chicago, IL 60637, USA}
\altaffiltext{24}{Kavli Institute for Cosmological Physics, University of Chicago, Chicago, IL 60637, USA}
\altaffiltext{25}{School of Physics and Center for Relativistic Astrophysics, Georgia Institute of Technology, 837 State Street NW, Atlanta, GA 30332-0430}
\altaffiltext{26}{Department of Physics, Anderson University, 1100 East 5th Street, Anderson, IN 46012}
\altaffiltext{27}{Physics Department, California Polytechnic State University, San Luis Obispo, CA 94307, USA}
\altaffiltext{28}{Department of Applied Physics and Instrumentation, Cork Institute of Technology, Bishopstown, Cork, Ireland}
\altaffiltext{29}{Now at Departament d’Astronomia i Meteorologia, Institut de Ciències del Cosmos, Universitat de Barcelona, IEEC-UB, Martí i Franquès 1, E-08028 Barcelona, Spain}
\altaffiltext{a}{\url{mccann@kicp.uchicago.edu}}
\altaffiltext{b}{\url{gtrichards@gatech.edu}}

\begin{abstract}
We present the results of 71.6~hours of observations of the Geminga
pulsar (PSR J0633+1746) with the VERITAS very-high-energy gamma-ray
telescope array. Data taken with VERITAS between November 2007 and
February 2013 were phase-folded using a Geminga pulsar timing solution
derived from data recorded by the XMM-\emph{Newton} and
\emph{Fermi}-LAT space telescopes. No significant pulsed emission
above 100~GeV is observed, and we report upper limits at the 95\%
confidence level on the integral flux above 135~GeV (spectral analysis
threshold) of 4.0$\times10^{-13}$~s$^{-1}$~cm$^{-2}$ and
1.7$\times10^{-13}$~s$^{-1}$~cm$^{-2}$ for the two principal peaks in
the emission profile. These upper limits, placed in context with
phase-resolved spectral energy distributions determined from five
years of data from the \emph{Fermi}-LAT, constrain possible hardening
of the Geminga pulsar emission spectra above $\sim$50~GeV.
\end{abstract}


\section{Introduction}
\setcounter{footnote}{0}
Following the completion of the Compton Gamma Ray Observatory (CGRO)
mission in 2000, seven gamma-ray pulsars were known to exist. A
combined total of 37 photons with energies exceeding 10~GeV were
observed from five of these pulsars by the EGRET instrument on-board
CGRO \citep{Thompson2005ApJS}. The \Fermi Large Area Telescope (LAT)
has now detected over 160 gamma-ray
pulsars\footnote{\url{https://confluence.slac.stanford.edu/display/GLAMCOG/Public+List+of+LAT-Detected+Gamma-Ray+Pulsars}}
(see \citealt{Caraveo2014ARA&A} for a review) and pulsar studies
presented in the \fermi-LAT catalog of sources above 10~GeV (1FHL)
have shown that 20 of these pulsars have \fermi-LAT detections above
10~GeV, with 12 also seen at energies above 25~GeV
\citep{SazParkinson2012,Ackermann2013ApJS}. One common feature
exhibited by all known gamma-ray pulsars is the shape of the spectral
energy distribution (SED), which can be described by a power law
followed by a spectral break occurring between 1 and 10 GeV
\citep{Abdo2013ApJS}. The 12 pulsars observed above 25~GeV are largely
drawn from the brightest of the \Fermi pulsars (F$_{\rm
  >100~MeV}>1.6\times10^{-7}$s$^{-1}$cm$^{-2}$) and thus are
sufficiently bright to be detected by \Fermi at these energies even as
their spectrum falls rapidly above the break. The most favored general
description of gamma-ray emission from pulsars in the \Fermi era
postulates that electrons are accelerated in the outer
magnetosphere. This acceleration is limited by the radiation of
synchrotron and curvature photons, leading to spectral
cut-offs. Outer-magnetospheric models (outer-gap or slot-gap models)
that implement this emission framework can, in general, reproduce the
pulsar light curves and SEDs measured by the \fermi-LAT.

Recently the Vela pulsar - the brightest known gamma-ray pulsar - has
been detected at energies above 30~GeV by
H.E.S.S.\footnote{\url{http://www.mpg.de/8287998/velar-pulsar}} and
above 50~GeV in the \fermi-LAT data
\citep{Leung2014arXiv1410.5208}. The Crab pulsar, however, remains the
only pulsar known to emit above 100~GeV. The power-law extension of
the Crab pulsar SED measured above the GeV break by VERITAS
\citep{Aliu2011Sci} and MAGIC \citep{Aleksi2011ApJ,Aleksi2012A} cannot
be easily explained by curvature emission from the outer magnetosphere
\citep{Aliu2011Sci,Lyutikov2012ApJb} unless the radius of curvature of
the magnetic field line is larger than the radius of the light
cylinder \citep{Bednarek2012}. Some recent models attribute the pulsed
very-high-energy (VHE; $E>$100~GeV) emission from the Crab pulsar to
inverse-Compton (IC) scattering originating in the outer magnetosphere
\citep{Lyutikov2012ApJb,Du2012ApJ,Lyutikov2012ApJ} or to IC scattering
from beyond the light cylinder
\citep{Aharonian2012Natur,Petri2012MNRAS}. The question of whether
Crab-pulsar-like non-exponentially-suppressed VHE spectra are common
in other gamma-ray pulsars, such as Geminga, has meaningful
implications for our understanding of the physics of particle
acceleration and emission from pulsars.

Located at a distance of $\sim$200~pc
\citep{Caraveo1996ApJ,Faherty2007Ap}, the Geminga pulsar is the
second-brightest steady GeV source in the gamma-ray sky and is the
original ``radio-quiet'' pulsar. It has a period of 237~ms, a
spin-down age of 3$\times10^{5}$~yr and a spin-down luminosity of
3.26$\times10^{34}$~erg~s$^{-1}$ \citep{Bignami1996ARA}. Originally
detected as an unidentified source of $\sim$100~MeV gamma-ray emission
by the SAS-2 and COS-B instruments
\citep{Fichtel1975ApJ,Bennett1977A&A}, its nature as a pulsar was
established following the detection of pulsed X-ray emission in data
recorded by the ROSAT satellite \citep{Halpern1992}. Reanalysis of the
COS-B and SAS-2 data, using the pulsar timing solution determined from
the ROSAT data, confirmed the MeV source to be a gamma-ray pulsar
\citep{Bignami1992Natur,Mattox1992ApJ}. Analysis of the available
EGRET data further confirmed the identification
\citep{Bertsch1992Natur}. The pulsed X-ray source is composed of
thermal radiation from hot-spots on the surface of the neutron star
and non-thermal magnetospheric emission
\citep{Caraveo2004Sci}. Detailed gamma-ray observations of the Geminga
pulsar have been made with the EGRET, \emph{AGILE} and \emph{Fermi}
space telescopes
\citep{Hasselwander1994,Fierro1998ApJ,Pellizzoni2009ApJ,FermiGem2010ApJ}.
Repeated radio searches have failed to find a radio-pulsar counterpart
\citep{Ramachandran1998AA,McLaughlin1999ApJ} while optical and UV
pulsations have been reported at the 3.5$\sigma$ and 5$\sigma$ level,
respectively \citep{Shearer1998AA,Kargaltsev2005ApJ}.

The Geminga pulsar has been a target for ground-based very-high-energy
gamma-ray detectors for over two decades. Limits on the pulsed
gamma-ray flux in the TeV regime at the $\sim$10\% Crab Nebula level
have been reported by the Whipple, HEGRA and PACT collaborations
\citep{Akerlof1993A,Aharonian1999A,Singh2009APh}, while the
Ootacamund, Durham and Crimean groups have reported weak evidence
($\sim$3$\sigma$ level) for pulsed emission at the $\sim$50-100\% Crab
Nebula level
\citep{Vishwanath1993A,Bowden1993JPhG,Neshpor2001Ast}. Given the far
higher sensitivity of current ground-based gamma-ray arrays, it seems
likely that these reported excesses are due to statistical
fluctuations. At multi-TeV energies, an unpulsed and spatially
extended source attributed to the Geminga pulsar wind nebula
\citep{Caraveo2003Sci} has been detected at the $\sim$20\% Crab Nebula
level by the Milagro water-Cherenkov telescope
\citep{Abdo2007ApJ,Abdo2009ApJ}. Weak evidence (2.2$\sigma$) for this
unpulsed source has also been reported at TeV energies by the Tibet
air-shower array \citep{Amenomori2010ApJ}.

The phase-averaged differential photon flux of the Geminga pulsar, as
measured by the \emph{Fermi}-LAT in the range 0.1-50~GeV, is well
described by a power law with an index of 1.3$\pm$0.01 at low
energies, followed by a spectral break at $\sim$2.5~GeV
\citep{FermiGem2010ApJ}. Above the break energy, a sub-exponential
cut-off in the spectrum is favored over a pure exponential or
super-exponential shape, as is commonly seen in the bright \Fermi
pulsars \citep{Abdo2013ApJS}. \cite{Lyutikov2012ApJ} argues that above
the spectral break, the spectrum can be described by a power law, a
behavior similar to what has been measured by VERITAS and MAGIC in the
Crab pulsar above the spectral break. Geminga is one of the 12 pulsars
detected above 25~GeV in the 1FHL with the highest-energy photon
attributed to the Geminga pulsar with a 95\% confidence level having
an energy of 33~GeV \citep{Ackermann2013ApJS}.

The remainder of this paper is structured in the following way. In
Section~2 we describe our observations of the Geminga pulsar with the
VERITAS gamma-ray telescope array and the \fermi-LAT data analyzed in
this work. In Section~3 we discuss the temporal analysis of the
\fermi-LAT data and describe the maximum-likelihood fitting procedures
used to derive spectral energy distributions. In this section we also
describe the VERITAS event processing and timing analysis. Section~4
details the results of the VERITAS and \fermi-LAT analyses, and in
Section~5 we provide some discussion and concluding remarks.

\section{Observations}
VERITAS is a ground-based gamma-ray telescope array located at the
Fred Lawrence Whipple Observatory at the base of Mount Hopkins in
southern Arizona \citep{Holder2006APh}. The array consists of four
imaging atmospheric-Cherenkov telescopes, each employing a tessellated
12~m Davies-Cotton reflector \citep{Davies1957SoEn} instrumented with
a photomultiplier-tube camera with a 3.5$\degree$ field of view.  The
VERITAS array is sensitive to gamma rays with energies between
$\sim$80~GeV and 30~TeV, with a nominal sensitivity sufficient to
detect, at the 5$\sigma$ level, a steady point-like source with 1\% of
the Crab Nebula flux in approximately 25~hrs. The VERITAS observations
of Geminga presented here were made under clear, moonless skies
between 2007 November and 2013 February. After data-quality selection,
the resulting observations span a total of 71.6~hours performed at an
average elevation of 72$\degree$. The data set spans three different
configurations of the VERITAS array: 2007 March to 2009 July, the
original array layout; 2009 August to 2012 July, the layout following
the relocation of one telescope; and 2012 August to present, following
the upgrade of the VERITAS cameras and trigger system (see
\citealt{Kieda2011ICRC} for further details). The data were acquired
in a mixture of \textit{ON} and \textit{wobble} (also known as
\textit{false source}) observation modes \citep{Fomin1994APh}.

The \Fermi Large Area Telescope is a space-based electron-positron
pair-conversion gamma-ray telescope composed of a silicon-strip
particle tracker interleaved with tungsten foil conversion layers
coupled to a cesium iodide calorimeter. It is sensitive to gamma rays
in the energy range between 20~MeV and 300~GeV. The LAT has a field of
view of $\sim$2.4~sr and operates primarily in an all-sky survey mode,
covering the entire sky approximately every three hours (see
\citealt{Atwood2009ApJ} for further details). The \fermi-LAT analysis
of the Geminga pulsar presented here uses 5.2~years of Pass-7
reprocessed photon data recorded by the \fermi-LAT between 2008 August
8th and 2013 October 18. The data were analyzed using the \Fermi
Science Tools version \texttt{v9r33p0-fssc-20140520}.

\section{Data Analysis}
\subsection{Fermi-LAT Analysis}
\begin{figure}
\centering \includegraphics[width=0.48\textwidth]{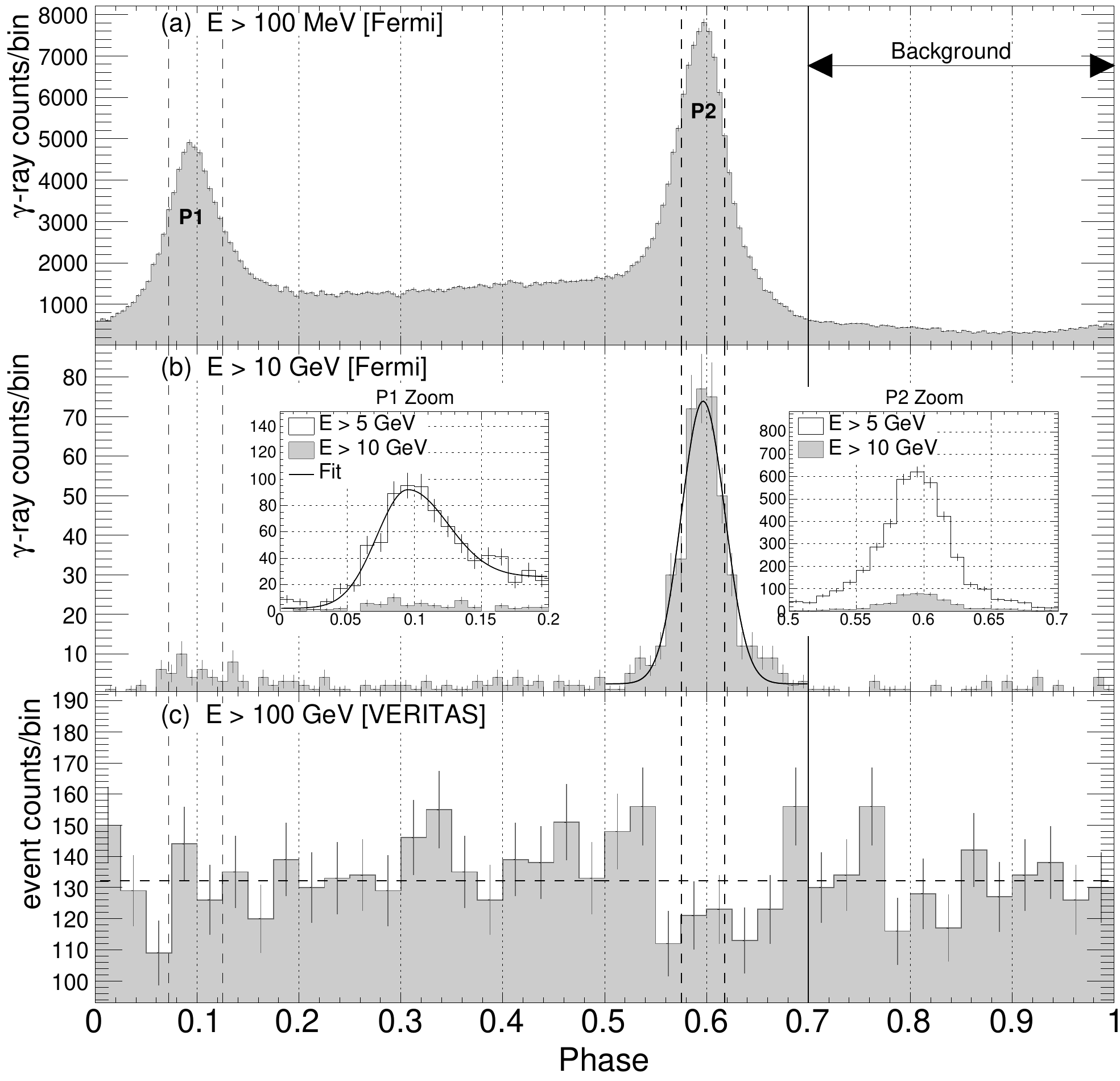}
\caption{The phase-folded light curve of the Geminga pulsar as
  measured by the \fermi-LAT. The \Fermi light curve contains all
  events that fell within a 2$\degree$ radius centered on the position
  of the Geminga pulsar. The energy-dependent evolution of the light
  curve is in clear agreement with the light curves presented in
  \cite{SazParkinson2012} and \cite{Ackermann2013ApJS}. The P1 and P2
  emission peaks were fitted with asymmetric Gaussian functions above
  5~GeV and 10~GeV, respectively. These fits, which are plotted as
  smooth black curves in panel (b), were used to define the signal
  regions for the P1 and P2 spectral analyses. These phase regions,
  [0.072 - 0.125] for P1 and [0.575 - 0.617] for P2, are indicated as
  vertical dashed lines. The background-event sample for the VERITAS
  analysis was selected from the phase range [0.7 - 1.0]. There is no
  evidence of pulsed emission above 100~GeV at any phase in the
  VERITAS data plotted in panel (c). }
\label{fig:prof}
\end{figure}

\begin{figure*}
\centering 
\includegraphics[width=0.99\textwidth]{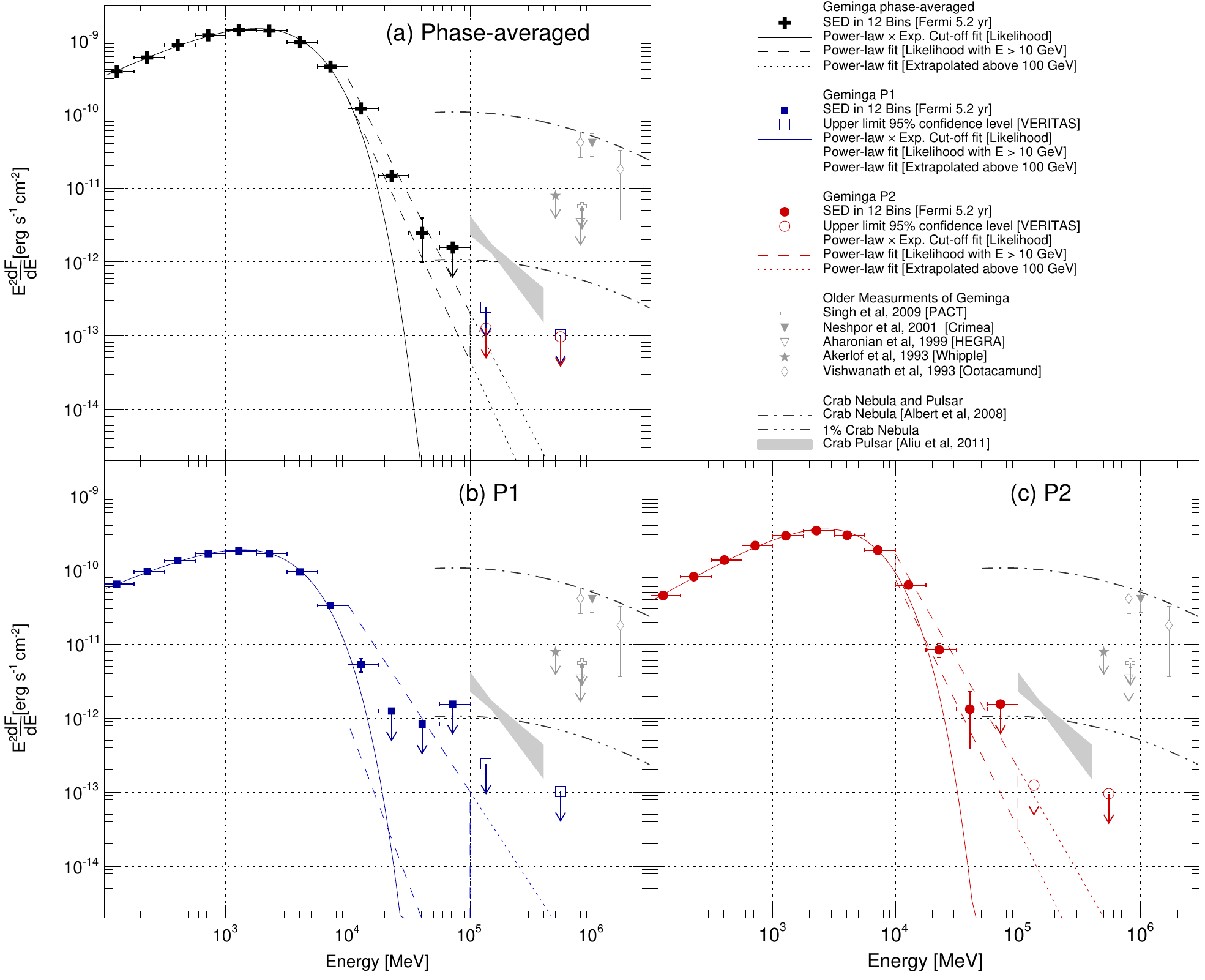}
\caption{Measured  SEDs   and  flux  upper  limits   for  the  Geminga
  pulsar. Measurements of  the Crab Nebula and pulsar  are plotted for
  comparison. The  Geminga limits and  fluxes shown for  PACT, Crimea,
  HEGRA, Whipple and Ootacamund  were derived from the integral values
  reported by those  experiments, assuming a power law  with index 2.5
  in each case.}
\label{fig:sed}
\end{figure*}

The \fermi-LAT analysis presented here follows the exact procedures
and analysis choices described in the second LAT catalog of gamma-ray
pulsars \citep{Abdo2013ApJS}.  \texttt{Source}-class photon events
within a 20$\degree$ region-of-interest (ROI) around the location of
the Geminga pulsar are selected, and time intervals when the edge of
the ROI extended beyond 100$\degree$ of the telescope zenith are
removed to prevent contamination by gamma rays from the Earth's
limb. Events are barycentered and phase-folded using the
\texttt{Tempo2} package \citep{Hobbs2006MNRAS} with the \Fermi
\texttt{Tempo2} plugin. The event times are folded using a timing
model for Geminga derived from \fermi-LAT data provided by Matthew
Kerr\footnote{\url{www.slac.stanford.edu/~kerrm/fermi_pulsar_timing/}}
\citep{Kerr2014}. The resulting pulsar light curve, which is dominated
by two emission peaks, labeled P1 and P2, connected by a ``bridge'' of
enhanced emission, is plotted in Figure~\ref{fig:prof}. The P1 and P2
peaks are fitted with asymmetric Gaussian functions above 5~GeV and
10~GeV, respectively. These energy cuts enable us to measure the width
of the emission peaks at high energies while maintaining good
statistics in each phase region. The $\pm1\sigma$ regions around each
peak (phases [0.072 - 0.125] for P1 and phases [0.575 - 0.617] for P2)
are then used as gates for phase-resolved \fermi-LAT spectra and as
signal regions for pulsed searches in the VERITAS data. To generate
the LAT spectra, binned maximum-likelihood analyses are performed in
12 logarithmically spaced energy bands between 100~MeV and 100~GeV. In
each energy band, a source model derived from the LAT 2-year
point-source catalog \citep{Nolan2012ApJS} is fitted to binned counts
maps in a 14$\degree\times$14$\degree$ region centered at the location
of the Geminga pulsar. The normalization of the galactic diffuse model
and the normalization of all sources within 4$\degree$ of Geminga are
allowed to float, while all other parameters are fixed to the 2-year
point-source catalog values. In each energy band, Geminga is modeled
as a point source with a power-law spectrum, floating normalization,
and a differential photon flux index fixed to the value 2. In
addition, binned likelihood analyses are performed across the entire
100~MeV to 100~GeV energy range with the same prescription as above,
with the differential photon flux of Geminga modeled as a power law
multiplied by an exponential cut-off:
\begin{linenomath}
\begin{equation}
\frac{dF}{dE} = A {\left(\frac{E}{E_{0}}\right)}^{-\Gamma}e^{-\left(\frac{E}{E_{\rm brk}}\right)}
\end{equation}
\end{linenomath}
where the normalization ($A$), index ($\Gamma$) and break energy
($E_{\rm brk}$) values allowed to float. The $E_{0}$ parameter is
fixed to the value 615.7~MeV, which is the decorrelation energy value
for Geminga reported in the LAT 2-year point-source catalog
\citep{Nolan2012ApJS}. Finally, and in order to probe a possible
power-law shape of the emission above the break, binned likelihood
analyses are performed between 10 and 100~GeV, modeling the Geminga
spectrum as a power law with floating normalization and index. The
SEDs for P1 and P2 derived from these likelihood analyses, where the
relevant cut on phase is applied to all events prior to performing the
likelihood fits, are plotted alongside the phase-averaged SED (where
no phase cut is applied) in Figure~\ref{fig:sed}. For each likelihood
fit, residual maps are generated between the measured counts map and
corresponding best-fit model map, and are found to show good agreement
between the data and model.

\subsection{VERITAS Analysis}
VERITAS data are passed through an analysis pipeline that reconstructs
the arrival direction and the energy of each gamma-ray candidate from
the Cherenkov images recorded by the telescopes. The images are
parametrized with the standard Hillas moment-analysis method
\citep{Hillas1985}. Event arrival directions and impact distances are
calculated from the stereoscopic images of the air showers from
multiple telescopes \citep{Hofmann1999APh}.  Background suppression,
i.e. cosmic ray rejection, and energy estimation are performed by
comparing measured event parameters to Monte Carlo gamma-ray
simulations with selection parameters combined in multidimensional
energy-dependent look-up tables \citep{Krawczynski06}. The optimal cut
values for a 1\% Crab Nebula strength source with a spectral index of
$\sim$4 are determined \textit{a priori} from the analysis of VERITAS
data on the BL Lac object PG~1553$+$113, which has this spectral index
value \citep{Aharonian2006A1553,Orr2011ICRC}.

After event selection, the event GPS times are converted to
barycentric dynamical time and phase-folded using
\texttt{Tempo2}. VERITAS events recorded prior to the launch of \Fermi
are folded using a timing solution derived from XMM-\emph{Newton}
observations of the Geminga pulsar \citep{GotthelfPC2014}. VERITAS
events recorded after the launch of \Fermi are folded using the
\fermi-LAT timing solution described in Section~3.1. The value of the
timing parameter \texttt{TZRMJD} in the XMM-\emph{Newton} model is
adjusted to ensure the definition of phase zero is consistent between
the two timing solutions.

VERITAS events that fall within the P1 and P2 phase gates are counted
as signal-plus-background events, with background-only events selected
from the phase region [0.7 - 1.0]\footnote{We note that, in this
  framework, VERITAS is not sensitive to the absolute flux level of
  the Geminga pulsar but to the difference in the flux level between
  the peak phase regions and the chosen background phase region. In
  contrast, the likelihood fitting employed in the \fermi-LAT analysis
  is sensitive to constant emission components. The Geminga pulsar
  flux above 100~MeV in the 0.7 to 1.0 phase range is $\lesssim$10\%
  of the flux level measured during the peak phases
  \citep{FermiGem2010ApJ}, thus any constant emission component is
  constrained to be at or below this level. Therefore any mismatch
  between the \fermi-LAT and VERITAS flux scales arising from the
  difference in background-estimation methods is within the systematic
  uncertainty on the absolute flux scale of both instruments.}. In
addition, an \textit{H}-Test for periodicity \citep{deJager1989AA} is
also performed on the VERITAS events. All steps in the VERITAS
analysis chain are cross checked and verified with an independent
analysis pipeline\footnote{Both analysis pipelines were used for the
  Crab pulsar data analysis presented in \cite{Aliu2011Sci}.}. Due to
the method of background estimation, the VERITAS analysis presented
here is not sensitive to unpulsed gamma-ray emission that might
originate from the pulsar magnetosphere or the pulsar wind nebula. An
analysis tuned for this type of unpulsed and possibly spatially
extended emission is ongoing and will be presented in a future
publication.

\section{Results}
\subsection{Light curve}
Our results from the analysis of 5.2 years of \fermi-LAT data (see
Figure~\ref{fig:prof}) are consistent with those previously reported
by \cite{FermiGem2010ApJ}, \cite{SazParkinson2012} and
\cite{Ackermann2013ApJS}. The light curve clearly evolves with energy,
with the P2 peak remaining visible at energies above 10~GeV, while the
P1 peak and the ``bridge'' are diminished. Above 100~GeV, there is no
evidence for pulsed emission. The VERITAS phase data plotted in
Figure~\ref{fig:prof}c have an \textit{H}-Test value of 1.8, which has
a probability of randomly occurring equal to 0.49. A $\chi^{2}$-fit of
the binned phase data for constant counts has a $\chi^{2}$/n.d.f value
of 45.95/39 and a fit probability of 0.2, indicating that the phase
distribution is entirely consistent with a random distribution.

\subsection{Spectrum}
The spectral analysis of 5.2 years of \fermi-LAT data (see
Figure~\ref{fig:sed}) are consistent with those previously reported by
the \fermi-LAT team \citep{FermiGem2010ApJ,Abdo2013ApJS}. The
phase-averaged SED, and the SEDs for P1 and P2, are all well described
by power laws with spectral breaks occurring between 1.8 and 2.8~GeV
(see Table~\ref{tab:FermiFit} for the best-fit values returned from
the maximum-likelihood analysis). Above the break, the SED data points
lie above the best-fit exponential cut-off function derived from the
likelihood analysis and appear more compatible with a pure power-law
function, as was previously noted by \cite{Lyutikov2012ApJ} (see the
dashed lines in Figure~\ref{fig:sed}). The last bin in the P1 SED with
a significant detection (likelihood test statistic $>$12) is between
10~GeV and 17~GeV, while the P2 and phase-averaged SEDs have
significant flux detected up to 56~GeV. We report upper limits on the
Geminga flux at the $\sim$1\% Crab Nebula level from the \fermi-LAT
data in the 50-100~GeV energy range.

In the VERITAS data above $\sim$100~GeV, the number of events falling
in the P1 and P2 signal regions is fully consistent with background
only (see Table ~\ref{tab:VTSsummary}). Using the method of
\cite{Helene1983}, the upper limit on the number of excess counts at
the 95\% confidence level is calculated. This upper limit divided by
the duration of the observation and effective area of VERITAS yields
the upper limit on the integral flux from the Geminga pulsar.  For the
integral flux upper limit calculation, a power law with a spectral
index of 3.8 is assumed, which is the same index value measured by
VERITAS in the Crab pulsar above 100 GeV
\protect\citep{Aliu2011Sci}. The resulting 95\% confidence level upper
limits are 4.0$\times10^{-13}$~s$^{-1}$~cm$^{-2}$ and
1.7$\times10^{-13}$~s$^{-1}$~cm$^{-2}$ on the integrated flux above
135~GeV\footnote{While this VERITAS analysis is sensitive to
  photons above $\sim$100~GeV in a search for pulsations, the
  threshold for spectral analysis is 135~GeV, and therefore upper
  limits are quoted above this energy.} for P1 and P2,
respectively. Above 550~GeV, the 95\% confidence level upper limits
are 5.1$\times10^{-14}$~s$^{-1}$~cm$^{-2}$ and
3.9$\times10^{-14}$~s$^{-1}$~cm$^{-2}$ for P1 and P2,
respectively. The corresponding energy fluxes, expressed in
erg~s$^{-1}$~cm$^{-2}$, are plotted in Figure~\ref{fig:sed}.

\begin{table*}
\centering
\begin{tabular}{c|ccc|cc}\hline
 Peak           &  \multicolumn{3}{c|}{ 100~MeV$<E<$100~GeV}                   &   \multicolumn{2}{c}{ 10~GeV$<E<$100 GeV}  \\\hline
                &  $A$                       & $\Gamma$        & $E_{\rm brk}$        &  $A$                       & $\Gamma$         \\
                &  [$\times10^{-10}$cm$^{-2}$  &                 &                 &  [$\times10^{-11}$cm$^{-2}$  &                  \\
                &  s$^{-1}$MeV$^{-1}$]         &                 & [GeV]           &  s$^{-1}$MeV$^{-1}$]         &                 \\\hline
P1              &   3.60$\pm$0.04            &  1.27$\pm$0.01  &  1.87$\pm$0.03  &   0.27$\pm$0.22            &  5.44$\pm$0.92  \\
P2              &   3.72$\pm$0.02            &  1.03$\pm$0.10  &  2.78$\pm$0.04  &   2.51$\pm$0.56            &  5.13$\pm$0.24  \\
Phase-averaged  &  22.60$\pm$0.07            &  1.23$\pm$0.01  &  2.33$\pm$0.02  &   5.83$\pm$1.02            &  5.37$\pm$0.19  \\\hline
\end{tabular}
\caption{Results from maximum-likelihood fits to the \fermi-LAT
  data. Between 100~MeV and 100~GeV the differential photon flux of
  Geminga was modeled as a power law multiplied by an exponential
  cut-off as defined in Equation~1. Between 10~GeV and 100~GeV, the
  differential photon flux of Geminga was modeled as a pure power law
  with the normalizing $E_{0}$ parameter fixed to 5~GeV. The quoted
  uncertainties are statistical only. The systematic uncertainty on
  the estimation of pulsar spectral values was studied by the
  \fermi-LAT collaboration in \cite{Abdo2013ApJS} and found to be, on
  average, 14\% for $\Gamma$ and 4\% for $E_{\rm brk}$. }
\label{tab:FermiFit}
\end{table*}

\begin{table*}
\centering
\begin{tabular}{c|cccccc}\hline
Peak        & \#Signal  & \#Background &  $\alpha$ & \#Scaled Background   & \#Excess   & Significance \\\hline
P1          &   284     &  1578       &  0.176    &  278.9  &   5.0      &  0.28$\sigma$ \\
P2          &   211     &  1578       &  0.141    &  223.7  & -12.7      & -0.80$\sigma$\\
P1$+$P2     &   495     &  1578       &  0.318    &  502.6  &  -7.6      & -0.29$\sigma$\\\hline
\end{tabular}
\caption{VERITAS event counts in the signal and background phase
  ranges. $\alpha$ is the ratio of the size of the signal phase gate
  to the background phase gate. The significance values were
  calculated using Equation~17 from \cite{LiMa1983ApJ}.}
\label{tab:VTSsummary}
\end{table*}

\section{Discussion and Conclusion}
Following a 71.6~hour exposure, we observe no significant pulsed
emission from the Geminga pulsar above 100~GeV.  The VERITAS 95\%
confidence level integral flux limits on the emission from the P1 and
P2 phase ranges limit any putative hard emission component above
135~GeV to be at or below the $\sim$0.25\% Crab Nebula level. These
limits represent the most constraining limits set to date on the
gamma-ray emission from the pulsar in this energy regime, surpassing
previous limits by over an order of magnitude. The spectral data
points derived from the analysis of 5.2 years of \fermi-LAT
observations are compatible with a power law up to the break energy,
but fall more slowly than what would be expected from a simple
exponential cut-off. It can be shown that the rounder, sub-exponential
shape, seen above the break in the phase-averaged SED, can be
reproduced by a superposition of several exponential cut-off functions
with different break energy \citep{FermiVela2010ApJ,
  Leung2014arXiv1410.5208}. Such a shape is expected in multizone
curvature-radiation models, when multiple acceleration regions with
different break energies combine to produce the observed
emission. Only at energies sufficiently above the maximum break energy
will the emission clearly fall exponentially.

Non-exponentially-suppressed emission above the GeV break energy,
expected in inverse-Compton emission pulsar models, has yet to be
conclusively observed with high significance in any pulsar other than
the Crab pulsar. In Geminga above 10~GeV, we see that pure power laws
with indices between 5.1 and 5.5 are compatible with the differential
photon flux points and predict a level of emission below the VERITAS
limits (see Table~\ref{tab:FermiFit} and Figure~\ref{fig:sed} for more
details). Similar results were found by \cite{Lyutikov2012ApJ}. A
cursory inspection of the Vela SED in \cite{Leung2014arXiv1410.5208}
suggests that a power law with an index of $\sim$2.4 ($\sim$4.4 for
the differential photon flux spectrum) is compatible with the data
points between 10 and 100~GeV, though the authors show the SED is well
fit by a multizone curvature emission model. 
Given this is the case for the two brightest gamma-ray pulsars, and
given the low fluxes from most pulsars above a few tens of
GeV\footnote{Only 4 of the 117 pulsars described in the second
  \fermi-LAT catalog of gamma-ray pulsars \citep{Abdo2013ApJS} have a
  measured flux point above 30~GeV with an average flux in the
  30-50~GeV range of 7.6$\times10^{-11}$~s$^{-1}$~cm$^{-2}$. For the
  remaining pulsars undetected in this energy range, the average 95\%
  confidence level flux upper limit is
  4.9$\times10^{-11}$~s$^{-1}$~cm$^{-2}$.}, we conclude that
power-law-type emission cannot be distinguished from the rounded
exponential cut-off shape expected in multizone curvature-emission
models with the available spectral data.

In the case of the Crab pulsar, and in several other cases where the
GeV break energy requires an acceleration efficiency close to or
exceeding unity \citep{Lyutikov2012ApJb}, canonical
curvature-radiation scenarios at the light cylinder are stressed. In
the case of Geminga, applying Formula~1 from \cite{Lyutikov2012ApJ}
and using the Geminga parameters from the ATNF Pulsar Catalog
\citep{Manchester2005AJ}, we find that the maximal break energy for
curvature radiation from the outer magnetosphere is $\epsilon_{br}$ =
2.53~GeV. The phase-averaged break-energy value reported here,
2.33$\pm$0.02~GeV, is consistent with this $\epsilon_{br}$. The P2
break energy, 2.78$\pm$0.04~GeV, does exceed the maximal curvature
break energy within the adopted outer-magnetospheric emission
framework.  We note, however, that $\epsilon_{br}$ is a function of
the assumed neutron-star radius and surface $B$-field strength to the
powers of $\nicefrac{9}{4}$ and $\nicefrac{3}{4}$,
respectively. Changes in either of these parameters at the 5-10\%
level bring the derived $\epsilon_{br}$ into agreement with our
measured value. However, the measured break energies in Geminga do
require the acceleration efficiency to approach unity at the light
cylinder in this radiation-reaction-limited curvature-emission
framework. This, in addition to the compatibility of the power law
shape with the high energy data, positions Geminga as a viable
candidate for inverse-Compton emission. Assuming the Cherenkov
Telescope Array (CTA) performs as expected \citep{Bernlohr2013APh},
future observations with CTA should be able to firmly detect the steep
power law extrapolated from the \fermi-LAT data at energies above
100~GeV in roughly one hundred hours.

\acknowledgments This research is supported by grants from the
U.S. Department of Energy Office of Science, the U.S. National Science
Foundation and the Smithsonian Institution, by NSERC in Canada, by
Science Foundation Ireland (SFI 10/RFP/AST2748) and by STFC in the
U.K. We acknowledge the excellent work of the technical support staff
at the Fred Lawrence Whipple Observatory and at the collaborating
institutions in the construction and operation of the instrument. The
VERITAS Collaboration is grateful to Trevor Weekes for his seminal
contributions and leadership in the field of VHE gamma-ray
astrophysics, which made this study possible. AMc is supported in part
by the Kavli Institute for Cosmological Physics at the University of
Chicago through grant NSF PHY-1125897 and an endowment from the Kavli
Foundation and its founder Fred Kavli.

\end{document}